\documentclass[onecollarge]{svjour2}
\usepackage[T1]{fontenc}
\usepackage[latin9]{inputenc}
\usepackage{textcomp}
\usepackage{url}
\usepackage{amsbsy}
\usepackage{amstext}
\usepackage{amssymb}
\usepackage{graphicx}
\usepackage{esint}
\usepackage[authoryear]{natbib}
\usepackage{epstopdf}
\usepackage{graphicx,subfigure}
\usepackage{subfig}
\RequirePackage{fix-cm}
\smartqed  
\journalname{Celestial Mechanics and Dynamical Astronomy}
\newcommand*\aap{Astron.~Astrophys. }
\newcommand*\aj{Astron.~J. }

\newcommand*\icarus{Icarus }
\newcommand*\mnras{Mon.~Not.~RAS }
\newcommand*\nat{Nature }
\newcommand*\planss{Planet.~Space~Sci. }
\newcommand*\apjl{Astrophys.~J.~Lett. }
\newcommand*\cmda{Cel.~Mech.~Dyn.~Astr. }
\newcommand*\celmec{Celest.~Mech. }
\newcommand*\sci{Science }
\newcommand*\mps{Meteor.~Plan.~Sci. }
\newcommand*\zamp{J.~Appl.~Math.~Phys. }
\newcommand*\primm{J.~Appl.~Math.~Mech. }

\begin{document}

\title{Capture Probability in the 3:1 Mean Motion Resonance with Jupiter}

\subtitle{An Application to the Vesta Family}

\author{H. A. Folonier \and F. Roig \and C. Beaug\'e}

\institute{H. A. Folonier \at \emph{Instituto de Astronomia Geof\'isica e Ci\^encias
Atmosf\'ericas, Universidade de S\~ao Paulo, Brasil} \\
 \email{folonier@usp.br} \\
 \and F. Roig \at \emph{Observat\'orio Nacional, Rio de Janeiro, Brasil}\\
 \email{froig@on.br}\\
 \and C. Beaug\'e \at \emph{Instituto de Astronom\'ia Te\'orica y Experimental
(IATE), Observatorio Astron\'omico de C\'ordoba, Universidad Nacional
de C\'ordoba, Argentina} \\
 \email{beauge@oac.uncor.edu}}

\titlerunning{Capture Probability in the 3:1 Resonance}

\authorrunning{H. A. Folonier et al.}

\date{Received: $\qquad\qquad$ / Accepted: $\qquad\qquad$}
\maketitle

\begin{abstract}
We study the capture and crossing probabilities into the 3:1 mean
motion resonance with Jupiter for a small asteroid that migrates
from the inner to the middle Main Belt under the action of the Yarkovsky
effect. We use an algebraic mapping of the averaged planar restricted
three-body problem based on the symplectic mapping of \citet{1993CeMDA..56..563H},
adding the secular variations of the orbit of Jupiter and non-symplectic
terms to simulate the migration. We found that, for fast migration
rates, the captures occur at discrete windows of initial eccentricities
whose specific locations depend on the initial resonant angles, 
indicating that the capture phenomenon is not probabilistic.
For slow migration rates, these windows become narrower and start
to accumulate at low eccentricities, generating a region of mutual
overlap where the capture probability tends to 100\%, in agreement
with the theoretical predictions for the adiabatic regime. Our simulations
allow to predict the capture probabilities in both the adiabatic and
non-adiabatic cases, in good agreement with results of \citet{1995CeMDA..61...97G}
and \citet{2006MNRAS.365.1367Q}. We apply our model to the case of
the Vesta asteroid family in the same context as \citet{2008Icar..194..125R},
and found results indicating that the high capture probability of
Vesta family members into the 3:1 mean motion resonance is basically
governed by the eccentricity of Jupiter and its secular variations.
\end{abstract}


\section{Introduction\label{intro}}

Asteroid taxonomy classify the asteroids in different types according
to the characteristics of their reflectance colors and/or spectra
(e.g. \citealt{1984PhDT.........3T}; \citealt{2002Icar..158..146B};
\citealt{2009Icar..202..160D}). V-type asteroids are a particular
class whose reflectance spectra have been recognized to be compatible
with the spectra of the basaltic achondritic meteorites (e.g. \citealt{1993Sci...260..186B};
\citealt{1998AMR....11..163H}; \citealt{2001M&PS...36..761B}). Currently,
the only known source for these asteroids is the Vesta family (e.g.
\citealt{1997M&PS...32..903M}; \citealt{2005Icar..174...54M}) a
group of asteroids in the inner Main Belt ($2.1<a<2.5$ AU) which
constitute the outcome of a collision that excavated the basaltic
surface of asteroid (4) Vesta more than 1 Gyr ago (e.g. \citealt{1970Sci...168.1445M};
\citealt{1997Sci...277.1492T}; \citealt{1997M&PS...32..965A}). Although
most of the V-type asteroids, the so called vestoids, are found within
the limits of the Vesta family ($2.22<a<2.47$ AU), several V-type
bodies are also found far away from the family outskirts (e.g. \citealt{1991Icar...89....1C};
\citealt{2000Sci...288.2033L}; \citealt{2008Icar..194..125R}; \citealt{2008ApJ...682L..57M};
\citealt{2009P&SS...57..229D}), which raises the question of how
these asteroids get to their current locations. Among the suggested
mechanisms, \citet{2008Icar..194..125R} show that some V-type asteroids
initially in the Vesta family would be able to cross the 3:1 mean
motion resonance (MMR) with Jupiter (hereafter J3:1) at 2.5 AU to
get to the middle Main Belt ($2.5<a<2.8$ AU). The crossing would
be driven by a slow migration of the orbital semimajor axis induced
by the thermal emission forces on the asteroid's surface, the so called
Yarkovsky effect (e.g. \citealt{2002aste.conf..395B}). Inspired by
this result, we address here the problem of resonance crossing/capture
in the case of the J3:1 MMR from a wider perspective, aiming to investigate
how this phenomenon happens and how the results of \citet{2008Icar..194..125R}
can be interpreted in the light of the general resonance crossing/capture
mechanism.

Different authors have proposed different dynamical mechanisms to
explain, at least partially, the presence of V-type asteroid in the
inner Main Belt beyond the domains of the Vesta family. \citet{2005A&A...441..819C}
showed that some V-type asteroids could have migrated from the Vesta
family to their current orbits due to the interplay between the Yarkovsky
effect and non-linear secular resonances. \citet{2008Icar..193...85N}
addressed a similar interplay, but with two-body and three-body MMRs.
Even the role close encounter with massive asteroids has also been
proposed as a mechanism (\citealt{2007A&A...465..315C}; \citealt{2012A&A...540A.118D}).
However, these mechanisms are not sufficient to account for all the
V-type asteroids found in the inner Main Belt.

To add to the puzzle, several V-type candidates have been recently
discovered in the middle Main Belt (e.g. \citealt{2006Icar..183..411R};
\citealt{2008Icar..198...77M}), and although most of them still lack
spectroscopic confirmation, they are strong photometric candidates.
Some of these bodies have moderate sizes, with diameters between 2-5
km, and their origin is still a matter of debate. For the time being,
the only reliable source of these asteroids should be the Vesta family.
However, in order to reach their present locations, these asteroids
should have crossed the J3:1 MMR which, at first hand, appears a near
impossible task due to the strong chaotic behavior that a
small asteroid temporarily trapped in this MMR would experiment. Actually,
it is well known that even in the simplest models, the J3:1 resonant
motion drives asteroids to high- and very high-eccentricity orbits
in less than a few tens of million years (\citealt{1982AJ.....87..577W};
\citealt{1991pscn.proc..177F}; \citealt{1996CeMDA..64...93F}). This
behavior allows the asteroids to cross the orbits of Mars and of the
Earth, thus being removed by close encounters with these planets.
\citet{2008Icar..194..125R}, used full N-body simulations, including
the perturbations of all the planets from Venus to Neptune, to find
that for asteroids with diameter of the order of 0.1-1.0 km there
is a small probability ($\sim3\%$) of crossing this resonance going
from the inner to the middle Main Belt. For larger bodies, the probability
would be even lower. In principle, the results of \citet{2008Icar..194..125R}
could only explain very few cases of V-type candidates in the middle
Main Belt.

The above scenario raises some questions like: Is the interaction
with the J3:1 MMR enough to explain other V-type candidates? How does
the resulting evolution depend on the asteroid's size? What dynamical
effects are relevant to the crossing/capture probability? Would other
planetary configurations lead to different results? These questions
shift the spotlight from the origin of the V-type asteroids to a more
fundamental issue: What is the capture/crossing probability in the
J3:1 MMR? And how does this vary in different dynamical models and
migration regimes?

The problem of resonance trapping has been approached by many authors.
\citet{1975PriMM..39R.621N} presented one of the first studies of
passages through a resonance separatrix with a slowly-varying parameter
(i.e. adiabatic regime). \citet{1979CeMec..19....3Y} calculated the
capture probability in the case of a simple pendulum, while \citet{1982CeMec..27....3H}
extended the study to the second fundamental model for first-order
resonances. The case of higher-order commensurabilities was undertaken
by \citet{1984CeMec..32..109L} and by \citet{1984CeMec..32..127B}.
Finally, \citet{1990Icar...87..249M} analyzed the capture in secondary
resonances including mutual inclination between the asteroid and the
perturbing planet.

All these works, however, deal with the adiabatic case in which the
migration timescale towards the resonance is much longer than the
libration period. For very small asteroids, however, the Yarkovsky
effect may lead to a non-adiabatic migration (e.g. \citealt{1998Icar..132..378F};
\citealt{2000Natur.407..606V}), and many of the classical predictions
by the above authors may not be valid. The problem of resonance capture/crossing
under a non-adiabatic regime is still little understood. \citet{1995CeMDA..61...97G}
studied the evolution of small particles migrating due to the Poynting-Robertson
drag. He found that the capture probability decreases for increasing
migration rates, especially for almost circular orbits. A similar
result was also found by \citet{2006MNRAS.365.1367Q}, who addressed
the case where the migrating body is the perturber. In particular,
this author introduces a simple semi-analytical model which allows
her to study the capture probability in a single MMR of any order
and also in the occurrence of a secondary resonance. Recently,
\citet{2011MNRAS.413..554M} used a Hamiltonian model to investigate
the capture probabilities in first and second order resonances considering
different scenarios like planet migration through a gas disk, through
a debris disk, and also dust migration under the Poynting-Robertson
drag. These authors found that resonant capture fails for high migration
rates, and has decreasing probability for higher eccentricities, although
for certain migration rates, capture probability peaks at a finite
eccentricity. They also found that more massive planets can capture
particles at higher eccentricities and migration rates.

In this work we focus on the behavior of convergent migration towards
the J3:1 MMR due to the Yarkovsky effect, both in the adiabatic and
non-adiabatic regimes. We are particularly interested in three key
issues: (i) how the capture probability changes with the migration
rate, (ii) the effects of different dynamical models, and (iii) an
application of these results to the Vesta family. We are also
interested in the behavior of the dynamical system for a wide range
of migration rates, and consequently will also discuss drifts that
correspond to meteoroid-size bodies. Since the orbital evolution of
such small particles is also influenced by other physical processes
(e.g. YORP, spin reorientations, etc), our results for this high non-adiabatic
regime should not be considered as accurate predictions, but solely
for theoretical completeness.

The paper is organized as follows. In Sect. \ref{model}, we present
our dynamical model and equations of motion. In Sect. \ref{mapping},
we introduce an algebraic mapping that allows us to follow the evolution
of a huge number of sets of initial conditions with less computational
cost. The probability of resonance capture in the adiabatic and non-adiabatic
cases is discussed in Sect. \ref{capture}. Numerical simulations
with our mapping in both adiabatic and non-adiabatic regimes are presented
in Sect. \ref{simula}. An application of our results to the present-day
distribution of V-type asteroids is given in Sect. \ref{compa}. Finally,
conclusions close the paper in Sect. \ref{conclu}.

\section{The Dynamical Model\label{model}}

Our analysis is based on a planar restricted three-body problem, consisting
of a massless particle (asteroid) orbiting a primary of mass $m_{0}$
(Sun) and perturbed by an exterior mass $m_{1}$ (Jupiter). We adopt
the usual Delaunay canonical variables: 
\begin{eqnarray}
L=\sqrt{\mu a}, & \ \ \ \ \ \ \ \ \  & \lambda=\textmd{mean longitude},\nonumber \\
L-G=\sqrt{\mu a}(1-\sqrt{1-e^{2}}), & \ \ \ \ \ \ \ \ \  & -\varpi=\textmd{longitude of pericenter},\label{delau}
\end{eqnarray}
where $\mu={\cal G}m_{0}$, ${\cal G}$ is the gravitational constant,
$a$ is the semimajor axis of the asteroid and $e$ its eccentricity.
The orbital elements of the perturbing mass (Jupiter) will be denoted
by a subscript $1$. The mean motions will be denoted by $n$ and
$n_{1}$, respectively. Orbital elements are assumed to be heliocentric.

In these variables, the Hamiltonian governing the orbital evolution
of the asteroid is given by the expression: 
\begin{equation}
{\cal H}=-{\displaystyle \frac{\mu^{2}}{2L^{2}}+n_{1}\Lambda-\frac{\mu_{1}}{a_{1}}\ R(L,L-G,\Lambda,\lambda,\varpi,\lambda_{1})},\label{hexact}
\end{equation}
where $R$ is the disturbing function, $\mu_{1}={\cal G}m_{1}$ and
$\Lambda$ is the canonical momentum associated to $\lambda_{1}=n_{1}t$.

Restricting the phase space to a vicinity of the J3:1 MMR, we can
expand $R$ in a Fourier-Poisson series (e.g. Laplace expansion) and
average over the short-period terms. Performing this averaging up
to first order in the masses, and retaining only terms up to second
order in the eccentricities, we can write the resonant Hamiltonian
as: 
\begin{eqnarray}
{\cal H} & = & -{\displaystyle \frac{\mu^{2}}{2L^{2}}+n_{1}\Lambda-}\nonumber \\
 &  & -{\displaystyle \frac{\mu_{1}}{a_{1}}\left[e^{2}A_{1}+ee_{1}A_{3}\cos(\varpi-\varpi_{1})+e^{2}A_{5}\cos(3\lambda_{1}-\lambda-2\varpi)\right.+}\nonumber \\
 &  & \ \ \ \ \ \ +\left.ee_{1}A_{6}\cos(3\lambda_{1}-\lambda-\varpi-\varpi_{1})+e_{1}^{2}A_{7}\cos(3\lambda_{1}-\lambda-2\varpi_{1})\right],\label{hreselem}
\end{eqnarray}
where, for simplicity, we have kept the same notation used for the
original Hamiltonian functions. In this expression $A_{i}$ are function
of the Laplace coefficients $b_{s}^{(j)}(a/a_{1})$ and are considered
constant. According to \citet{1999ssd..book.....M} their values at
the J3:1 MMR $(a/a_{1}=0.48075)$ are: 
\begin{eqnarray*}
A_{1} & = & 0.142097,\ \ \ \ \ A_{3}=-0.165406,\ \ \ \ \ A_{5}=0.598100,\\
A_{6} & = & -2.21124,\ \ \ \ \ A_{7}=0.362954.
\end{eqnarray*}

Since we are interested in the motion around the J3:1 commensurability,
we may transform our variables to their resonant counterparts (e.g.
\citealt{1983CeMec..30..197H}): 
\begin{eqnarray}
\sigma=\frac{1}{2}(3\lambda_{1}-\lambda)-\varpi, & \ \ \ \ \ \ \ \ \  & S=(L-G),\nonumber \\
-\nu=\frac{1}{2}(3\lambda_{1}-\lambda)-\varpi_{1}, & \ \ \ \ \ \ \ \ \  & N=(L-G)-L-\Lambda,\\
Q=\frac{1}{2}(\lambda_{1}-\lambda), & \ \ \ \ \ \ \ \ \  & \bar{\Lambda}=-\Lambda-3L.\nonumber 
\end{eqnarray}
As the averaged Hamiltonian does not depend explicitly on $Q$, its
momentum $\bar{\Lambda}$ is a constant of motion. Without any loss
in generality we can take its value equal to zero, from which we can
attain that $\Lambda=-3L$. In this manner, the system is reduced
to 2 degrees of freedom and the remaining canonical momenta acquire
the form: 
\begin{eqnarray}
S & = & (L-G)=\sqrt{\mu a}\big(1-\sqrt{1-e^{2}}\big),\nonumber \\
N & = & (L-G)+2L=\sqrt{\mu a}\big(3-\sqrt{1-e^{2}}\big).\label{actions}
\end{eqnarray}

Finally, since we are only retaining terms in $R$ up to the second
order in the eccentricities, we can write $e$ as a function of both
$S$ and $N$ up to the same order as: 
\begin{equation}
e\approx2\sqrt{\frac{S}{N}}.
\end{equation}
Writing the resonant Hamiltonian Eq. (\ref{hreselem}) in terms of
the resonant canonical variables, we obtain, up to the second order
in the eccentricities: 
\begin{eqnarray}
{\cal H} & = & -\frac{2\mu^{2}}{(N-S)^{2}}-\frac{3}{2}n_{1}(N-S)-\frac{\mu_{1}}{a_{1}}\left[4\frac{S}{N}\left(A_{1}+A_{5}\cos(2\sigma)\right)\right.+\nonumber \\
 &  & +\left.2e_{1}\sqrt{\frac{S}{N}}\left(A_{3}\cos(\sigma+\nu)+A_{6}\cos(\sigma-\nu)\right)+e_{1}^{2}A_{7}\cos(2\nu)\right].\label{hrescanon}
\end{eqnarray}
The resulting equations of motion can be explicitly written as: 
\begin{eqnarray}
{\displaystyle \frac{dS}{dt}} & = & -\frac{\mu_{1}}{a_{1}}\frac{8S}{N}A_{5}\sin(2\sigma)-\frac{\mu_{1}}{a_{1}}2e_{1}\sqrt{\frac{S}{N}}\left[A_{3}\sin(\sigma+\nu)+A_{6}\sin(\sigma-\nu)\right]\nonumber \\
{\displaystyle \frac{dN}{dt}} & = & -\frac{\mu_{1}}{a_{1}}2e_{1}^{2}A_{7}\sin(2\nu)-\frac{\mu_{1}}{a_{1}}2e_{1}\sqrt{\frac{S}{N}}\left[A_{3}\sin(\sigma+\nu)-A_{6}\sin(\sigma-\nu)\right]\nonumber \\
{\displaystyle \frac{d\sigma}{dt}} & = & -\frac{4\mu^{2}}{(N-S)^{3}}+\frac{3}{2}n_{1}-\frac{\mu_{1}}{a_{1}}\frac{4}{N}\left[A_{1}+A_{5}\cos(2\sigma)\right]-\nonumber \\
 &  & -\frac{\mu_{1}}{a_{1}}\frac{e_{1}}{\sqrt{S\, N}}\left[A_{3}\cos(\sigma+\nu)+A_{6}\cos(\sigma-\nu)\right]\nonumber \\
{\displaystyle \frac{d\nu}{dt}} & = & \frac{4\mu^{2}}{(N-S)^{3}}-\frac{3}{2}n_{1}+\frac{\mu_{1}}{a_{1}}\frac{4S}{N^{2}}\left[A_{1}+A_{5}\cos(2\sigma)\right]+\nonumber \\
 &  & +\frac{\mu_{1}}{a_{1}}e_{1}\sqrt{\frac{S}{N^{3}}}\left[A_{3}\cos(\sigma+\nu)+A_{6}\cos(\sigma-\nu)\right].\label{hameqs}
\end{eqnarray}

\section{The Algebraic Mapping\label{mapping}}

To solve these variational equations, we implemented an algebraic
mapping based on the symplectic mapping introduced by Hadjidemetriou
(\citeyear{1986ZaMP...37..776H}, \citeyear{1991pscn.proc..157H},
\citeyear{1993CeMDA..56..563H}) and by \citet{1996CeMDA..65..421F},
to which we added a non-symplectic term (e.g. \citealt{1996CeMDA..65..407C})
simulating the migration due to the Yarkovsky effect.

The classical Hadjidemetriou's mapping is a variation of the twist
map, tailored to preserve the fixed points of the original Hamiltonian
as well as their stability indices. If we write the averaged Hamiltonian
Eq. (\ref{hrescanon}) as: 
\begin{equation}
{\cal H}(\boldsymbol{I},\boldsymbol{\theta})={\cal H}_{0}(\boldsymbol{I})+\mu_{1}{\cal H}_{1}(\boldsymbol{I},\boldsymbol{\theta}),
\end{equation}
where $(\boldsymbol{I},\boldsymbol{\theta})\equiv(S,N,\sigma,\nu)$,
${\cal H}_{0}$ is the integrable part and ${\cal H}_{1}$ is the
disturbing function, the mapping at the $i$-th step is given by a
canonical transformation $(\boldsymbol{I}_{i},\boldsymbol{\theta}_{i})\rightarrow(\boldsymbol{I}_{i+1},\boldsymbol{\theta}_{i+1})$
with a Jacobi-type generating function: 
\begin{equation}
\mathcal{S}(\boldsymbol{I}_{i+1},\boldsymbol{\theta}_{i})=\boldsymbol{I}_{i+1}\cdot\boldsymbol{\theta}_{i}+\tau{\cal H}(\boldsymbol{I}_{i+1},\boldsymbol{\theta}_{i}).
\end{equation}
From this expression, the implicit form of the mapping is given by:
\begin{eqnarray}
\boldsymbol{I}_{i} & = & {\displaystyle \frac{\partial\mathcal{S}}{\partial\boldsymbol{\theta}_{i}}=\boldsymbol{I}_{i+1}+\tau{\displaystyle \frac{\partial{\cal H}(\boldsymbol{I}_{i+1},\boldsymbol{\theta}_{i})}{\partial\boldsymbol{\theta}_{i}}}},\nonumber \\
\boldsymbol{\theta}_{i+1} & = & {\displaystyle \frac{\partial\mathcal{S}}{\partial\boldsymbol{I}_{i+1}}=\boldsymbol{\theta}_{i}+\tau{\displaystyle \frac{\partial{\cal H}(\boldsymbol{I}_{i+1},\boldsymbol{\theta}_{i})}{\partial\boldsymbol{I}_{i+1}}.}}\label{map}
\end{eqnarray}

The time-step $\tau$ of the mapping must be set to the period of
the synodic angle $Q$ over which the Hamiltonian was averaged; see
\citet{1993CeMDA..56..563H} for a detailed construction. This is
$\tau=2\pi/n_{1}\approx11.86$ yr, that corresponds to Jupiter's orbital
period. Substituting the partial derivatives by the left-hand part
of Eqs. (\ref{hameqs}), leads to the expression of the mapping for
the J3:1 MMR. It is worth noting that the two equations for the actions
$\boldsymbol{I}\equiv(S,N)$ are given in implicit form and must be
solved iteratively before solving the two equations for the angles
$\boldsymbol{\theta}\equiv(\sigma,\nu)$.

\subsection{Adding the Yarkovsky Effect}

The next step is to add to the algebraic mapping a non-conservative
term mimicking the Yarkovsky effect acting on small asteroids. The
main consequence of the Yarkovsky effect is a secular drift of the
semimajor axis of the asteroid, but no changes in either the eccentricity
or the angles. In other words, the total time variation of the semimajor
axis can be expressed as the sum of two components: 
\begin{equation}
\dot{a}=\dot{a}_{G}+\dot{a}_{Y},
\end{equation}
where $\dot{a}_{G}$ is due to the purely gravitational perturbations,
while $\dot{a}_{Y}$ is the variation due to the non-conservative
term.

According to \citet{1999A&A...344..362V}, the rate of change of the
semimajor axis due to the diurnal version of the Yarkovsky effect
can be given approximately by: 
\begin{equation}
\dot{a}_{Y}=\kappa_{d}\,\frac{1\,\mathrm{km}}{D}\,\cos\epsilon,\label{basalt}
\end{equation}
where $D$ is the diameter of the asteroid in km, $\epsilon$ is the
obliquity of the spin axis with respect to the orbital plane, and
$\kappa_{d}$ is a constant that depends on several physical and thermal
parameters of the asteroid, like the albedo, the surface thermal conductivity,
the surface and bulk densities, the surface emissivity, and the rotational
period. Assuming values of these quantities typical of the vestoids,
we have $\kappa=2.5\times10^{-10}$ AU/yr (e.g. \citealt{2008Icar..193...85N}).
We note that, if $\cos\epsilon>0$, the asteroid increases its semimajor
axis and pulls away from the Sun, while if $\cos\epsilon<0$, it suffers
an orbital decay. Since we wish to study convergent migration towards
the J3:1 MMR from smaller values of the semimajor axis, we assume
$\cos\epsilon=1$ in order to maximize the effect%
\footnote{There is also a seasonal version of the effect which produces a drift
of the form: 
\[
-\kappa_{s}\,\frac{1\,\mathrm{km}}{D}\,\sin^{2}\epsilon,
\]
but for the problem in hand, the constant $\kappa_{s}\ll\kappa_{d}$,
thus the seasonal effect can be disregarded.%
}. For the rest of this paper, we will apply Eq. (\ref{basalt})
as an approximate link between a given migration rate and the corresponding
asteroid diameter. Since the value of $\kappa_{d}$ has been estimated
from large vestoids, it is not clear that its value will remain
invariant for smaller asteroids and meteoroids. However,
our aim is not to give quantitatively accurate values of the body
diameters, but to present illustrative quantities.

From Eqs. (\ref{actions}), we can see that both canonical momenta
depend on the semimajor axis, so both will be affected by the Yarkovsky
effect, that is: 
\begin{equation}
\dot{\boldsymbol{I}}=\frac{\partial\boldsymbol{I}}{\partial a}\dot{a}+\frac{\partial\boldsymbol{I}}{\partial e}\dot{e}=\left(\frac{\partial\boldsymbol{I}}{\partial a}\dot{a}_{G}+\frac{\partial\boldsymbol{I}}{\partial e}\dot{e}\right)+\frac{\partial\boldsymbol{I}}{\partial a}\dot{a}_{Y},
\end{equation}
and taking into account that $\boldsymbol{I}=\sqrt{\mu a}\,\Phi(e)$,
we have: 
\begin{equation}
\dot{\boldsymbol{I}}=\dot{\boldsymbol{I}}_{G}+\frac{2\mu\boldsymbol{I}}{(N-S)^{2}}\dot{a}_{Y},
\end{equation}
where $\dot{\boldsymbol{I}}_{G}$ is given by the first two Eqs. (\ref{hameqs}).

Introducing this expression into the mapping, we obtain: 
\begin{eqnarray}
\boldsymbol{I}_{i+1} & = & {\displaystyle \boldsymbol{I}_{i}-\tau\mu_{1}{\displaystyle \frac{\partial{\cal H}_{1}(\boldsymbol{I}_{i+1},\boldsymbol{\theta}_{i})}{\partial\boldsymbol{\theta}_{i}}+\tau\dot{a}_{Y}{\displaystyle \frac{2\mu\boldsymbol{I}_{i+1}}{(N_{i+1}-S_{i+1})^{2}}}}},\nonumber \\
\boldsymbol{\theta}_{i+1} & = & {\displaystyle \boldsymbol{\theta}_{i}+\tau{\displaystyle \frac{\partial{\cal H}_{0}(\boldsymbol{I}_{i+1})}{\partial\boldsymbol{I}_{i+1}}.}+\tau\mu_{1}{\displaystyle \frac{\partial{\cal H}_{1}(\boldsymbol{I}_{i+1},\boldsymbol{\theta}_{i})}{\partial\boldsymbol{I}_{i+1}}.}}\label{mapy}
\end{eqnarray}
It is worth noting that the addition of the Yarkovsky term breaks
the symplectic structure of the original mapping at $\boldsymbol{I}_{i+1}$
or at $\boldsymbol{I}_{i}$. This is not a problem since the Yarkovsky
effect acts as a dissipation, thus we should not expect the conservation
of the Hamiltonian $\mathcal{H}$. Note also that, if $\dot{a}_{Y}>\mu_{1}$,
the non-conservative term may become as important as the gravitational
perturbation itself. In our simulations, we always consider drift
values such that $\dot{a}_{Y}<\mu_{1}$.

\subsection{Adding Long Period Terms of the Perturber's Orbit\label{longper}}

Following \citet{1996CeMDA..65..421F} and \citet{1999P&SS...47..653R},
the mapping can be further improved by adding the secular perturbations
on the orbit of Jupiter. From the classical planetary theory, we have
that the secular change of the eccentricity $e_{1}$ and perihelion
longitude $\varpi_{1}$ is given by a sum of harmonic terms: 
\begin{equation}
e_{1}\exp\mathrm{i}\varpi_{1}=\sum_{k}G_{k}\exp[\mathrm{i}(\gamma_{k}t+\phi_{k})]\qquad\qquad\mathrm{i}=\sqrt{-1}.\label{khjup}
\end{equation}
The values for the amplitudes $G_{k}$, frequencies $\gamma_{k}$,
and initial phases $\phi_{k}$ of the harmonics are listed in Table
\ref{tab1}, and they were adopted from the synthetic secular theory
LONGSTOP 1B (\citealt{1989A&A...210..313N}). In particular, we considered
only the principal harmonics, i.e. those with amplitude $G_{k}\geq10^{-4}$.
The only exceptions are the harmonics $g_{8}$ and $2g_{5}-g_{6}$,
which are not excluded because $g_{8}$ is one of the fundamental
frequencies of the planetary theory, and $2g_{5}-g_{6}$ is the retrograde
frequency of Jupiter's perihelion with the longest period. Equation
(\ref{khjup}) is directly introduced at each time step of the mapping,
$t=i\tau$, through $e_{1}$ and $\nu$ in the right-hand part of
Eqs. (\ref{hameqs}) and (\ref{mapy}).

\begin{table}
\caption{Frequencies $\gamma_{k}$, initial phases $\phi_{k}$ and amplitudes
$G_{k}$ for the secular variation of Jupiter's orbit. Initial phases
are given at JD 2440400.5. As usual, $g_{i},s_{i}$ represent the
fundamental frequencies of the perihelia and nodes, respectively,
of the planets (5 for Jupiter, 6 for Saturn, and so on).}
\label{tab1}
\begin{tabular}{crrrr}
\hline 
Harmonic term  & $\gamma_{k}[\mathrm{"/yr}]\quad$  & $\phi_{k}[{}^{\circ}]\;$  & $G_{k}\qquad\qquad$  & Period {[}yr{]}$\quad$\tabularnewline
 &  &  &  & \tabularnewline
$g_{5}$  & $4.25749319$  & $27.0005$  & $4.41872\times10^{-2}$  & $3.0440\times10^{5}$\tabularnewline
$g_{6}$  & $28.24552984$  & $124.1994$  & $-1.57002\times10^{-2}$  & $4.5883\times10^{4}$\tabularnewline
$g_{7}$  & $3.08675577$  & $117.0516$  & $1.8139\times10^{-3}$  & $4.1986\times10^{5}$\tabularnewline
$g_{8}$  & $0.67255084$  & $70.7508$  & $5.80\times10^{-5}$  & $1.9270\times10^{6}$\tabularnewline
$g_{5}+g_{6}-g_{7}$  &  &  & $-1.936\times10^{-4}$  & $4.4057\times10^{4}$\tabularnewline
$-g_{5}+g_{6}+g_{7}$  &  &  & $1.982\times10^{-4}$  & $4.7867\times10^{4}$\tabularnewline
$-g_{5}+2g_{6}$  &  &  & $-5.735\times10^{-4}$  & $2.4812\times10^{4}$\tabularnewline
$2g_{5}-g_{6}$  &  &  & $1.42\times10^{-5}$  & $-6.5685\times10^{4}$\tabularnewline
$g_{5}-s_{6}+s_{7}$  &  &  & $1.104\times10^{-4}$  & $4.6940\times10^{4}$\tabularnewline
$-g_{5}+2g_{6}+s_{6}-s_{7}$  &  &  & $-1.226\times10^{-4}$  & $4.4873\times10^{4}$\tabularnewline
\end{tabular}
\end{table}

\subsection{Comparison of the Mapping with the Full Hamiltonian Equations\label{mapxhex}}

In order to verify the validity of our mapping model, we performed
a series of numerical simulations with the mapping and compared the
results to those obtained from a direct N-body integration of the
full Hamiltonian (Eq. \ref{hexact}), using a Bulirsch-Stoer integrator.

We chose 500 initial conditions in the $a-e$ plane over a line segment
parallel to the left branch separatrix of the J3:1 MMR in the circular
problem ($e_{1}=0$). Taking the initial angular variables as $\theta=2\sigma=\pi$
and $\Delta\varpi=\sigma+\nu=\pi$, this line of initial conditions
follows the equation ${\displaystyle a=2.49-e/10}$, with $0.01\leq e\leq0.4$.
The orbital elements of Jupiter were fixed to $a_{1}=5.202545$ AU,
$e_{1}=0$ or 0.048, $\varpi_{1}=0$, and $\lambda_{1}=0$.

We adopted three different values for the Yarkovsky drift rate $\dot{a}_{Y}$,
corresponding to a very fast migration ($\dot{a}_{Y}=5\times10^{-5}$
AU/yr; $D=0.5$ cm), an intermediate migration ($\dot{a}_{Y}=5\times10^{-6}$
AU/yr; $D=5$ cm), and a slow migration ($\dot{a}_{Y}=5\times10^{-7}$
AU/yr; $D=50$ cm). The total integration time span was $T=2.0\times10^{3}$,
$2.0\times10^{4}$, and $2.0\times10^{5}$ yr, respectively for each
drift rate. The time step of the N-body code was automatically adjusted
to match a tolerance of $10^{-11}$ in the precision of the solution.

\begin{figure}[t]
\centering{}\includegraphics[clip,width=1\textwidth]{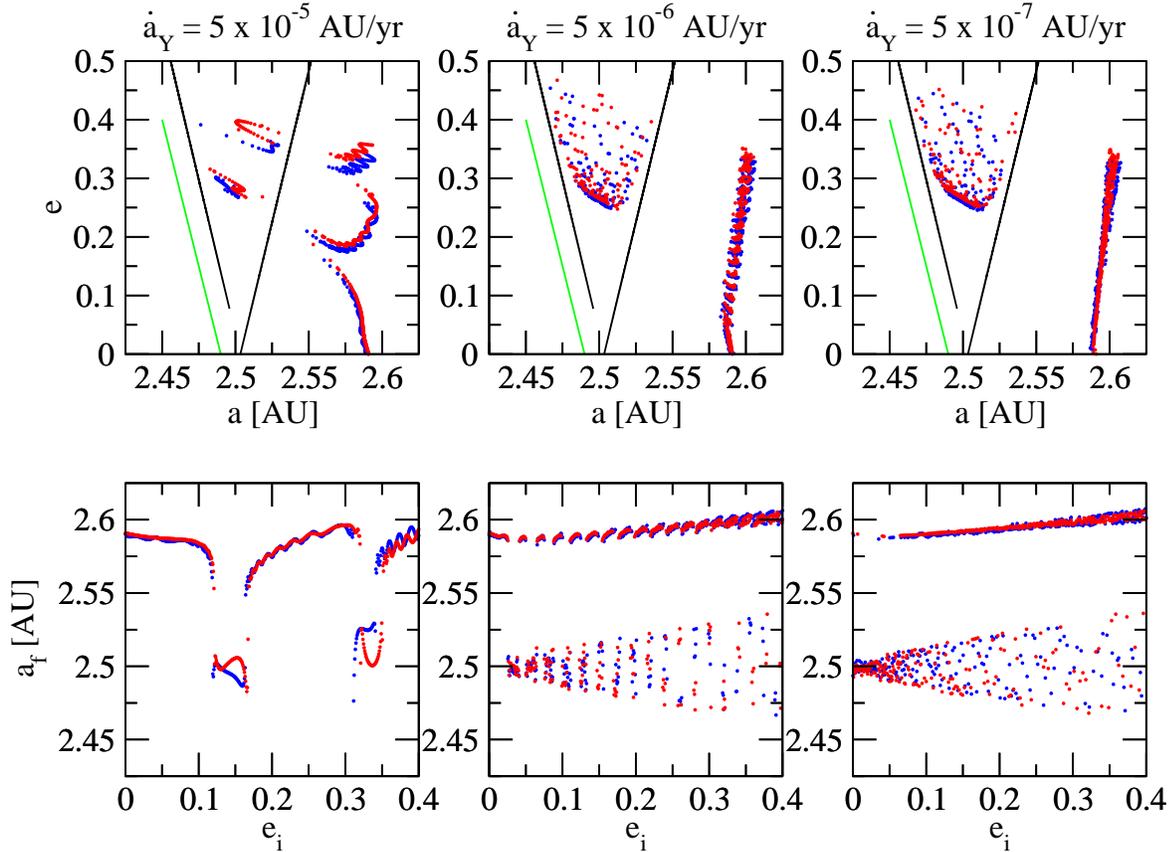}
\caption{Comparison between the algebraic mapping (red dots) and the integration
of the full N-body equations (blue dots) in the circular case ($e_{1}=0$).
The initial angles are set to $\theta=\Delta\varpi=\pi$. The migration
rate decreases from left to right. \emph{Top panels:} Final values
of the semimajor axes and eccentricities. The green line represents
the initial conditions, which are the same for both simulations. The
separatrix of the resonance are shown for reference (black lines).
\emph{Bottom panels:} Final semimajor axes vs. initial eccentricities,
where the structure of captures and crossing windows is appreciable
(see also Fig. \ref{fig1a}).}
\label{fig1}
\end{figure}

In Fig. \ref{fig1}, we show the results of this comparison for the
circular case. The black V-shaped lines in the upper row panels are
the separatrix of the circular problem, and the green line is the
set of initial conditions. The red dots are the final conditions (at
$t=T$) of the simulations obtained with the mapping, while the blue
dots were obtained with the N-body integration. In the lower row panels,
we show the final semimajor axes as a function of the initial eccentricity.
The leftmost panels correspond to the fastest migration rate, while
the rightmost ones correspond to the slowest rate.

\begin{figure}[th!]
\centering{}\includegraphics[width=0.9\textwidth]{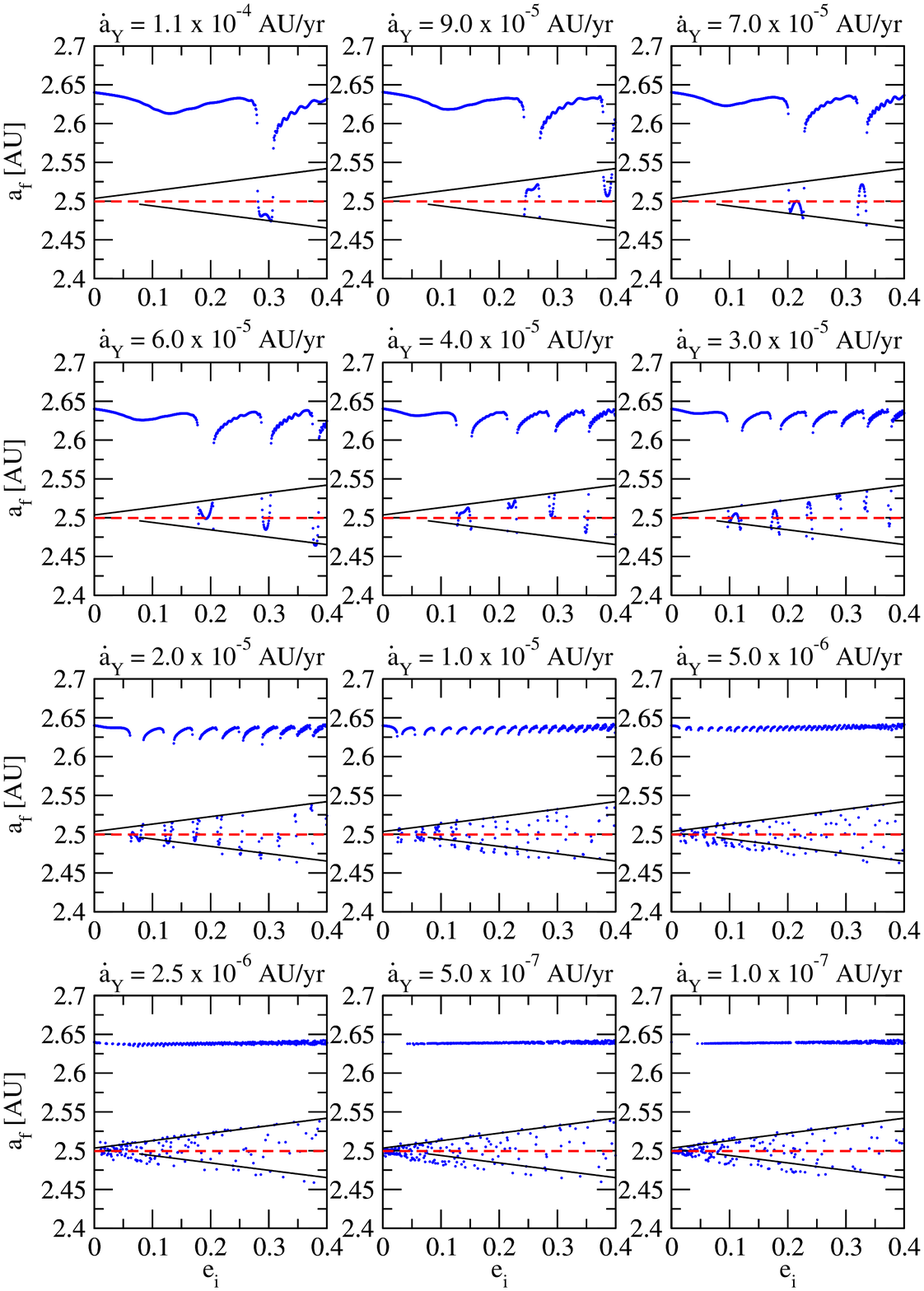}
\caption{Final semimajor axis vs. initial eccentricity obtained with the mapping
equations for the circular problem ($e_{1}=0$), showing the structure
of capture and crossing windows at fast migration rates. The initial
angles are set to $\theta=\Delta\varpi=0$. The migration rate decreases
from left to right and from top to bottom. The separatrix (black curves)
and center (red dashed) of the J3:1 MMR are indicated. This behavior
is also observed in the elliptic case, and also with the numerical
integration of the full Hamiltonian.}
\label{fig1a}
\end{figure}

We found a very good agreement between the mapping results and the
N-body integrations, both showing the same structures of crossings
and captures. For the fastest migration rates, the resonance captures
appear to occur in certain discrete ``windows'' of initial eccentricities
(lower left plot in Fig. \ref{fig1}). In other words, there exists
certain intervals of the initial eccentricities for which capture
always occurs, and other values for which a resonance crossing is
guaranteed. \emph{Thus, the outcome of the resonance passage is not
probabilistic, but well defined and deterministic}. 

This can be seen more clearly along a more detailed range of fast
migration rates, as shown in Fig. \ref{fig1a}. As the migration rate
decreases, the number of capture windows increases and the windows
become narrower. New windows start to appear at high eccentricities,
while those already present shift to the lower values of $e$. It
is worth noting that in Fig. \ref{fig1a} the initial angles were
fixed to $\theta=2\sigma=0$ and $\Delta\varpi=\sigma+\nu=0$. By
comparing the panels in Figs. \ref{fig1} and \ref{fig1a} corresponding
to the same migration rate ($5\times10^{-5}$ AU/yr), it becomes evident
that the precise location of the capture windows strongly depends
on the initial angles $\theta$ and $\Delta\varpi$.

As we approach the adiabatic regime (Fig. \ref{fig1}, bottom middle
and right), the capture windows tend to accumulate and overlap in
the region $0\leq e<e_{c}\approx0.04$, leading to certain capture
for all initial conditions in this eccentricity range. However, for
$e>e_{c}$ the windows do not overlap but continue to reduce in width
while their mutual separation decreases, tending to zero in the adiabatic
limit and leading to a probabilistic treatment of the outcome of any
initial condition. As we will show later, this behavior is in agreement
with the resonance capture analytical model in the adiabatic regime.

Although these results assumed that the perturber is in a circular
orbit, both the capture/crossing windows and their main characteristics
are also observed in more complete dynamical models, such as the elliptic
case and models including secular perturbations in the perturbing
planet. Finally, it is worth noting that the computation time could
be about 1,000 times faster with the mapping compared to an N-body
integration. This allows to perform a huge amount of simulations with
the mapping to obtain a statistically significant result.

\subsection{Capture dependence on the initial angles\label{dependence}}

In this section, we investigate how the capture process depends
on the initial angles, aiming to provide an explanation for the capture
windows observed in Figs. \ref{fig1} and \ref{fig1a}. We considered
a grid of test orbits with $a=2.45$ AU and $e=0.2$, and
initial angles $\theta$ and $\Delta\varpi$ varying between $0$ and
$360^{\circ}$. These test orbits were integrated until either $a>2.56$ AU
(crossing) or $e>0.5$ (capture).

\begin{figure}
\centering{}\includegraphics*[width=0.9\textwidth]{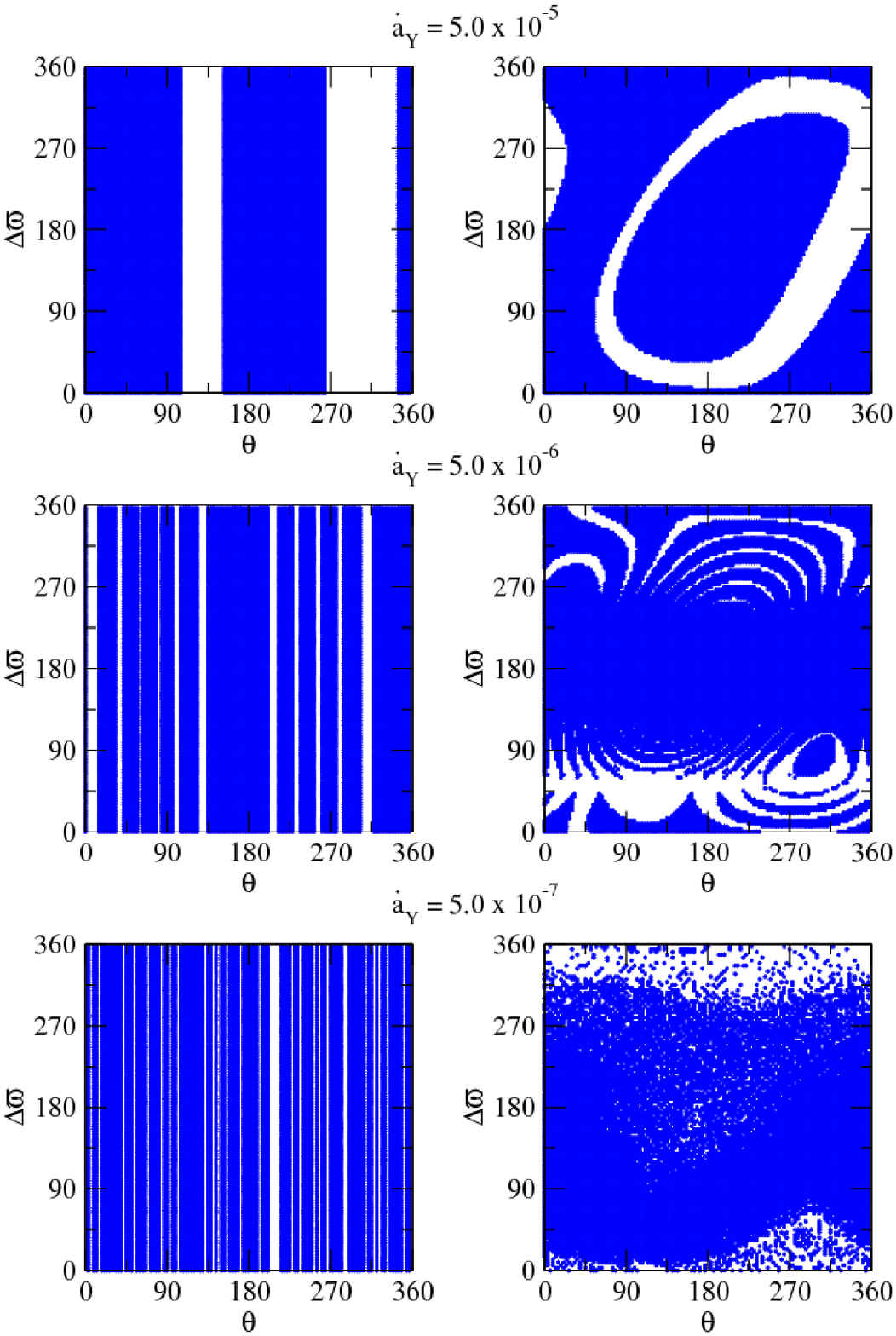}
\caption{Capture regions (in white) depending on the initial angles
$\theta=2\sigma$ and $\Delta\varpi=\sigma+\nu$ for a test orbit
with initial $a=2.45$ AU and $e=0.2$. The left panels correspond
to the circular model ($e_{1}=0$), while the right panels correspond
to the secular elliptic model (see Sect. \ref{longper}). The considered
drift rates are indicated above the panels. Angles are in degrees.}
\label{fig3new}
\end{figure}

The results are shown in Fig. \ref{fig3new}, for three different
Yarkovsky drift rates (the same shown in Fig. \ref{fig1}) and two
different models: the circular problem (left panels) and the secular
elliptic problem where the eccentricity of Jupiter varies with time
(right panels). The final fate of the orbits is identified by a blue
color for crossings and white for captures. It is worth noting that
the results for the pure elliptic model ($e_{1}\neq0$ fixed) are
very similar to those of the secular elliptic model, being actually
indistinguishable for the fastest drift rates.

Looking at the results for the circular model, which does
not depend on $\Delta\varpi$, it is clear that the occurrences of
captures are strongly dependent on the initial values of the resonant
angle. This behavior can be explained by assuming that the capture
takes place only if the orbit reaches the separatrix at, or very close
to, the unstable (saddle) equilibrium point, which correspond to $\theta=0$
in the J3:1 MMR. Since $e$ remains almost constant before reaching the 
resonance region, the rate of circulation $\dot{\theta}$ depends primarily
on the initial semimajor axis and all the test orbits reach the
separatrix at approximately the same time $\tau$. Therefore, for an initial angle $\theta_{0}$, 
the evolution of $\theta$ upon reaching the separatrix is approximately given
by $\theta-\theta_{0}=\int_{0}^{\tau}\dot{\theta}(\dot{a}_{Y}t)\, dt.$
Since the right-hand member is almost equal for all the orbits, only a limited 
range of $\theta_{0}$ values could lead to $\theta \sim 0$ at the separatrix.

For the fast migration rates, $\tau\sim 300$ yr is smaller than the circulation 
period ($\dot{\theta}\lesssim 1,000$ yr), and the orbits reach the separatrix before they can 
complete a full circulation of the resonant angle. Thus, we should expect a limited 
number of capture windows. On the other hand, for the slowest migration rates,
the orbits are able to make several circulations before reaching the
separatrix ($\tau\sim 30,000$ yr), providing a larger number of capture
windows. This produces the structure shown in the left panels of 
Fig. \ref{fig3new}.

In the case of the elliptic models, the captures are still
expected to occur through the saddle point $\theta=0$, but the evolution
is coupled with $\Delta\varpi$, which produces the complex patterns
observed in the right panels of Fig. \ref{fig3new}.

\section{The Resonance Capture Probability\label{capture}}

\subsection{Adiabatic Case}

\begin{figure}[ht]
\centering{}
\includegraphics*[width=0.49\textwidth]{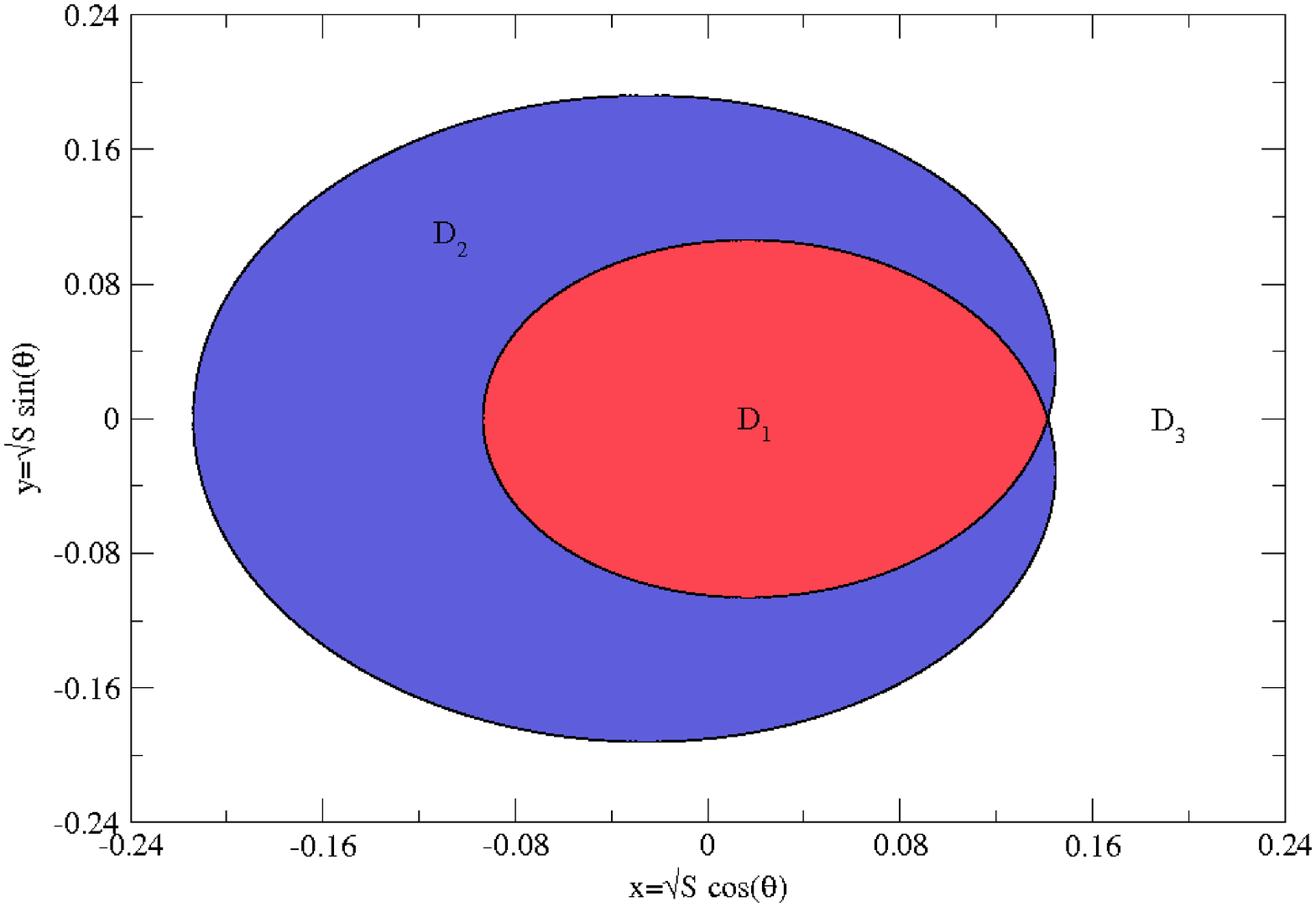}
\includegraphics*[width=0.49\textwidth]{folonier_fig04b.eps}
\caption{\emph{Left:} Schematic representation in the plane $x=\sqrt{S}\cos(\theta)$, 
$y=\sqrt{S}\sin(\theta)$ of a resonant, one degree of freedom, system 
with action-angle variables $(S,\theta)$. Region $D_{1}$ corresponds 
to a regime of inner circulation of $\theta$; region $D_{2}$ corresponds 
to a libration around $\theta=\pi$; and region $D_{3}$ corresponds 
to an outer circulation. \emph{Right:} Capture probability as a function 
of the eccentricity in the adiabatic case (black curve) and in three 
non adiabatic cases (blue, red and green curves). The green curve 
corresponds to the fastest migration rate. For slower migration rates, 
the location of the maximum shifts to smaller eccentricities, and 
the maximum probability grows until reaching the adiabatic case.}
\label{fig3}
\end{figure}

Figure \ref{fig3} (left) presents a schematic view of the phase space
of a pendulum-type one degree of freedom dynamical system. We can
define three domains: an inner circulation domain $D_{1}$ (in red),
a libration domain $D_{2}$ (in blue), and an outer circulation domain
$D_{3}$ (in white). We are interested in assessing the probability
$P_{ij}$ for an initial condition in region $D_{i}$ to pass onto
region $D_{j}$. Let $J_{i}$ be the area of region $D_{i}$, such
that $J_{3}=J_{1}+J_{2}$. In the absence of migration, these areas
remain constant and only depend on the value of the momentum $N$,
i.e. $J_{i}=J_{i}(N)$.

Following \citet{1982CeMec..27....3H}, when a very slow migration
is considered (adiabatic regime), the probability $P_{3i}$ is given
in a first approximation by: 
\begin{equation}
P_{3i}=\frac{\partial J_{i}/\partial N}{\partial J_{3}/\partial N}.\label{proba}
\end{equation}
This equation states that the capture probability is directly proportional
to that ratio of the speed at which the areas $D_{i}$ and $D_{3}$
change due to the migration. It is straightforward to show that $P_{31}=1-P_{32}$,
since: 
\begin{equation}
P_{31}+P_{32}={\displaystyle \frac{\partial J_{1}/\partial N}{\partial J_{3}/\partial N}+{\displaystyle \frac{\partial J_{2}/\partial N}{\partial J_{3}/\partial N}={\displaystyle \frac{\partial(J_{1}+J_{2})/\partial N}{\partial J_{3}/\partial N}=1,}}}
\end{equation}

To apply the above equation to the J3:1 MMR, we consider the Hamiltonian
of the restricted circular three-body problem (Eq. (\ref{hrescanon});
$e_{1}=0$) and, for each value of $N$, we numerically compute the
values of $J_{1}$ and $J_{3}$, as well as their derivatives. A condition
passing from region $D_{3}$ to $D_{2}$ constitutes a capture, while
a condition passing from region $D_{3}$ to $D_{1}$ constitutes a
crossing. Therefore, the capture probability is: 
\begin{equation}
P_{cap}=P_{32},
\end{equation}
while the crossing probability is: 
\begin{equation}
P_{cross}=P_{31}=1-P_{cap}.
\end{equation}

To find $P_{cap}$ as a function of $N$, we need to compute the ratio
between the derivatives in Eq. (\ref{proba}). Since the Yarkovsky
drift included in our model do not modify the eccentricity, we can
replace $N$ by the value of $e$ that the orbit has when it reaches
the separatrix of the resonance. This allows us to study the capture
probability directly as a function of $e$, as shown in the black
curve of Fig. \ref{fig3} (right).

These calculations indicate that, for eccentricities $0\leq e<e_{c}\approx0.05$,
the capture probability is 100 \%, while for $e>e_{c}$ it decays
exponentially. In other words, for an adiabatic migration, any asteroid
that enters the J3:1 MMR with $e<e_{c}$ will be captured. The capture
probability reduces to less than 15 \% for eccentricities larger than
0.4.

\subsection{Non-Adiabatic Case}

The capture probability in the non-adiabatic case is much more complex
and little is known about its behavior. Probably the first general
study was performed by \citet{1995CeMDA..61...97G}, who focused on
the dynamics of small particles entering a MMR domain induced by the
Poynting-Robertson effect. Using N-body numerical simulations, he
found that, independently of the initial eccentricity, the capture
probability decreased for faster migration rates. He also found that,
in the non-adiabatic regime, the capture probability is zero for circular
orbits, then increases up to certain maximum value at a given $e_{max}$,
and then decreases again asymptotically to zero for higher eccentricities.
The values of $e_{max}$ and the corresponding maximum of $P_{cap}$
depend on the migration rate: the faster the migration, the higher
the $e_{max}$ and the smaller the $P_{cap}$. This behavior is shown
in Fig. \ref{fig3} (right), which reproduces the curves estimated
by \citet{1995CeMDA..61...97G}.

More recently, the non-adiabatic case has also been studied by \citet{2006MNRAS.365.1367Q}.
A detailed discussion of her results in comparison with ours will
be treated in Sect. \ref{intpcap}.

\section{Mapping Simulations\label{simula}}

\subsection{Capture Probability}

In order to estimate the theoretical capture probability (in both
the adiabatic and non-adiabatic regimes) with our mapping simulations,
we considered three different models, depending on Jupiter's orbit:
the circular model ($e_{1}=0$, hereafter CM), the elliptic model
($e_{1}\neq0$ fixed, hereafter EM), and the secular elliptic model
(hereafter SEM) in which Jupiter's orbit feels the secular perturbations
of the other major planets according to Eq. (\ref{khjup}).

For the CM, we chose 1000 equispaced initial conditions over the line
${\displaystyle a=2.49-\frac{e}{10}}$, with $0.01\leq e\leq0.4$,
parallel to the left branch of the separatrix of the (circular) J3:1
MMR in the $a-e$ plane. For each of these initial conditions, we
considered 36 equispaced values of $\theta=2\sigma$ between 0 and
$2\pi$. For the EM and SEM, we chose 100 initial values of $(a,e)$
along the same lines, and for each of these we took 18$\times$18
equispaced values of $\theta$ and $\Delta\varpi=\sigma+\nu$, both
between $0$ and $2\pi$. It is worth noting that the proximity of
these initial conditions to the resonance separatrix implies that
the values of $e$ upon reaching the separatrix are almost the same
as their initial values. Therefore, from now on, we will analyze the
capture probability directly as a function of the initial eccentricity
of the orbits.

The initial conditions of Jupiter's orbit were $a_{1}=5.202545$ AU,
$e_{1}=0$ for the CM, $e_{1}=0.048$ for the EM and SEM, $\lambda_{1}=0$
and $\varpi_{1}=0$. Each set of initial conditions was integrated
with the mapping using 12 different migration rates $\dot{a}_{Y}$.
In Table \ref{tab2}, we list these values together with the corresponding
diameters according to Eq. (\ref{basalt}). The Table also shows the
total integration time span in each case, in units of mapping iterations,
where each iteration has period $\tau$.

\begin{table}
\caption{Assumed migration rates $\dot{a}_{Y}$ and their associated diameters
$D$. The last three columns give the number of iteration steps used
in the simulations of the different models. Each step corresponds
to a period of $\approx11.8$ yr.}
\label{tab2}
\begin{tabular}{ccccc}
\hline 
$\dot{a}_{Y}$ {[}AU/yr{]}  & $D$ {[}m{]}  & CM  & EM  & SEM \tabularnewline
 &  &  &  & \tabularnewline
$5.0\times10^{-4}$  & $5.0\times10^{-4}$  & 35  & 90  & 90 \tabularnewline
$2.5\times10^{-4}$  & $1.0\times10^{-3}$  & 90  & 100  & 100 \tabularnewline
$1.0\times10^{-4}$  & $2.5\times10^{-3}$  & 150  & 150  & 150 \tabularnewline
$5.0\times10^{-5}$  & $5.0\times10^{-3}$  & 200  & 250  & 250 \tabularnewline
$2.5\times10^{-5}$  & $0.01$  & 350  & 500  & 500 \tabularnewline
$1.0\times10^{-5}$  & $0.025$  & 700  & 1,000  & 1,000 \tabularnewline
$5.0\times10^{-6}$  & $0.05$  & 1,300  & 3,500  & 3,500 \tabularnewline
$2.5\times10^{-6}$  & $0.1$  & 2,500  & 10,000  & 15,000 \tabularnewline
$1.0\times10^{-6}$  & $0.25$  & 6,000  & 25,000  & 30,000 \tabularnewline
$5.0\times10^{-7}$  & $0.5$  & 12,500  & 60,000  & 75,000 \tabularnewline
$2.5\times10^{-7}$  & $1.0$  & 25,000  & 100,000  & 125,000 \tabularnewline
$1.0\times10^{-7}$  & $2.5$  & 60,000  & 300,000  & 300,000 \tabularnewline
\end{tabular}
\end{table}

\begin{figure}[th]
\centering{}\includegraphics*[clip,width=0.9\textwidth]{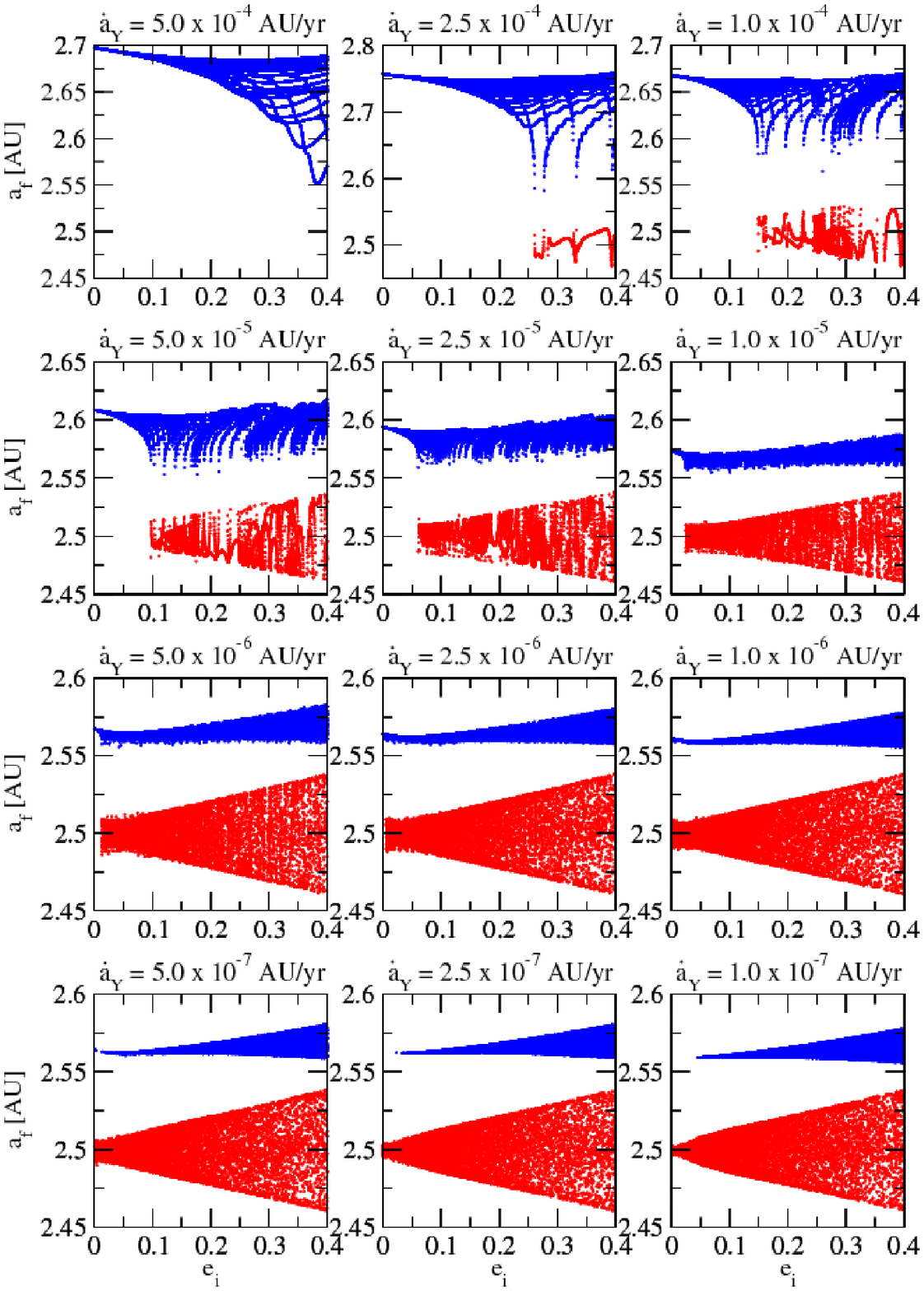}
\caption{Final semimajor axes vs. initial eccentricity obtained from the numerical
simulations with the mapping assuming Jupiter in a circular orbit
(model CM). Red dots represent captures while blue dots represent
crossings. The migration rate $\dot{a}_{Y}$ decreases from left to
right and from top to bottom.}
\label{fig7}
\end{figure}

\begin{figure}[th]
\centering{}\includegraphics*[clip,width=0.9\textwidth]{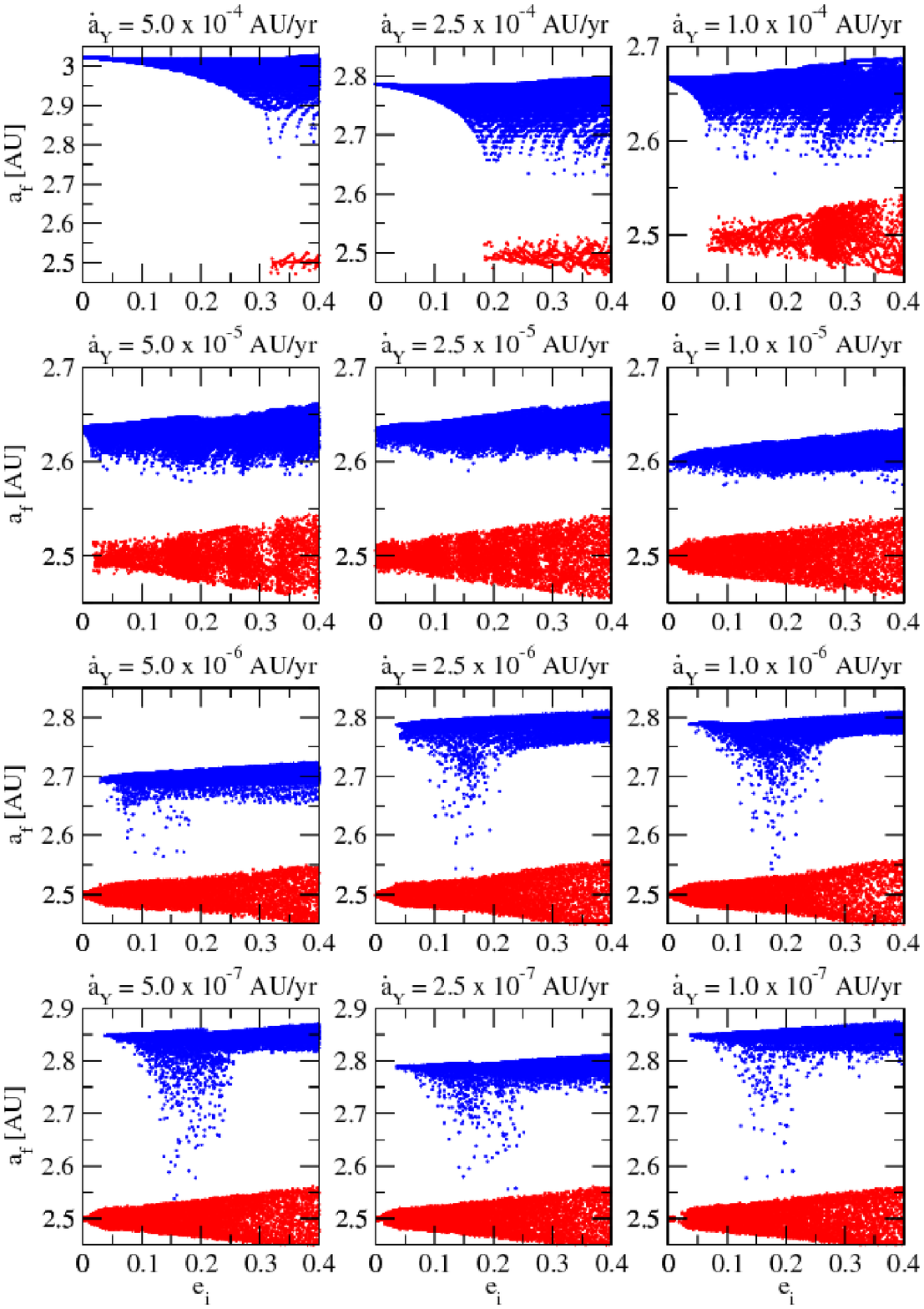}
\caption{Same as Fig. \ref{fig7} but for Jupiter in a fixed elliptic orbit
(EM). Note the tail of ``delayed'' crossings that appears between
$0.1<e<0.2$ for the slowest migration rates (last six panels).}
\label{fig8}
\end{figure}

In Figs. \ref{fig7} and \ref{fig8}, we show the final semimajor
axes as a function of the initial eccentricities for models CM and
EM, respectively. The results of the secular elliptic model (SEM)
shows no significant differences with those adopting a fixed orbit
for the perturber, and are not shown. Each frame corresponds to a
different migration rate, starting from the fastest (upper left) to
the slowest (bottom right). We assume that an asteroid has crossed
the resonance if its final semimajor axis is larger than a critical
value, which varies from case to case but was always of the order
of $a_{c}\approx2.55$ AU. The capture conditions are represented
in red, while the crossings are shown in blue. Since we have now considered
an ensemble of initial angles $(\theta,\Delta\varpi)$, the windows
associated to captures/crossing become blurred and there is no longer
a predetermined outcome, even for fast migration rates.

\begin{figure}[th]
\centering{}\includegraphics*[clip,width=0.9\textwidth]{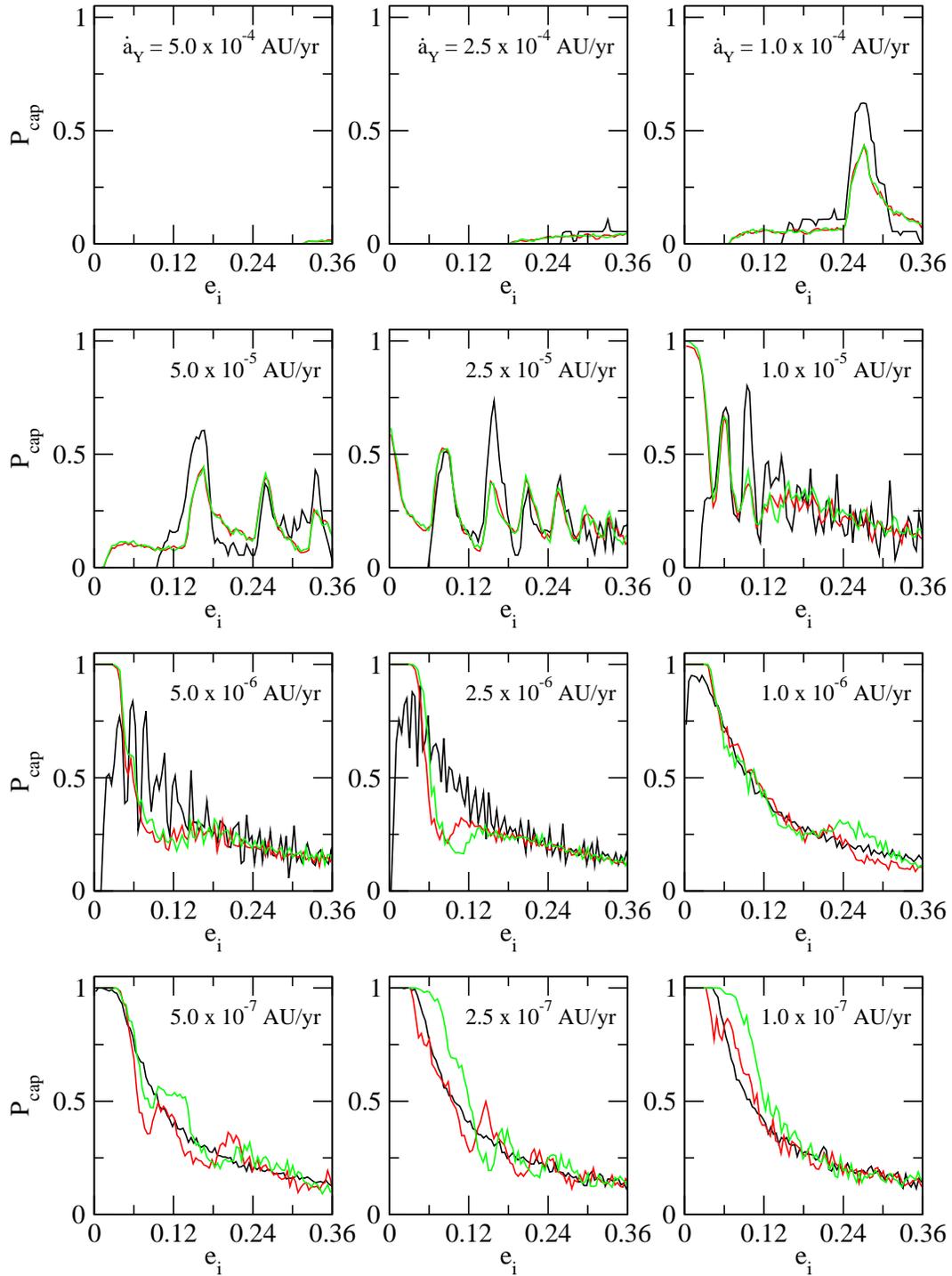}
\caption{Capture probability as a function of the initial eccentricity computed
from the simulations shown in Figs. \ref{fig7} (CM) and \ref{fig8}
(EM), plus results from the secular elliptic model (SEM). The black
curve corresponds to the CM, red to the EM, and green to the SEM.
The migration rate decreases from left to right and from top to bottom.
For each initial eccentricity, the computed probability is ``averaged''
over the resonant angles $\theta,\Delta\varpi$.}
\label{fig10}
\end{figure}

The capture probability was computed by counting the number of captures
that occur for each initial eccentricity and dividing it by the total
number of initial conditions at that eccentricity. Rigorously speaking,
the capture probability depends on the migration rate $\dot{a}_{Y}$,
the initial value of $e$, and the initial angles $\theta$ and $\Delta\varpi$,
as shown in Sect. \ref{mapxhex}. Since, for each eccentricity, the
initial angles are varied between 0 and $2\pi$, our counting method
is equivalent to estimate an integrated probability: 
\begin{equation}
\langle P_{cap}\rangle_{\theta,\Delta\varpi}=P_{cap}(e,\dot{a}_{Y})={\displaystyle \frac{1}{(2\pi)^{2}}\int_{0}^{2\pi}\int_{0}^{2\pi}P_{cap}(e,\dot{a}_{Y},\theta,\Delta\varpi)\ d\theta d\Delta\varpi.}
\end{equation}
The values of $P_{cap}(e,\dot{a}_{Y})$ are shown in Fig. \ref{fig10}
for the CM (black curve), the EM (red curve), and the SEM (green curve).

At the fastest migration rate, all the initial conditions were able
to cross the resonance in the CM, but in the EM and SEM a few captures
were registered for large initial eccentricities ($e>0.3$). Captures
in the CM start to happen at migration rates slower than $\dot{a}_{Y}=2.5\times10^{-4}$
AU/yr ($D=0.1$ cm); however, in the EM and SEM captures already occur
at lower eccentricities ($e>0.2$) at the same migration rates. As
expected, the capture windows appear initially at high eccentricities
and move toward smaller $e$ as the migration rate slows down (cf.
Sect. \ref{mapxhex}).

At a migration rate of $\dot{a}_{Y}=1.0\times10^{-5}$ AU/yr ($D=2.5$
cm), the capture probability in the CM resembles the curve of a non-adiabatic
capture (see Fig. \ref{fig3}), but with a lot of overlapped noise.
This noise is introduced by the discretization of the capture windows,
and it is not observed in \citet{1995CeMDA..61...97G} since this
author performed a smoothing of his capture curves. The capture probability
starts with a maximum at large eccentricities, and drifts towards
smaller eccentricities taking higher values as the migration rate
becomes slower. In the EM and SEM, and for migration rates slower
than $\dot{a}_{Y}=5.0\times10^{-6}$ AU/yr ($D=5$ cm), we observe
a tail of final conditions that cross the resonance at low initial
eccentricities (blue dots) This tail is formed by orbits that spent
much more time inside the resonance than the remaining orbits before
jumping it. This delay is probably due to the interaction of the orbits
with secular and secondary resonances inside the MMR; that is why
the same effect is not observed in the CM. At slower migration rates,
this delay effect is observed for $0.05<e<0.25$.

For both the EM and SEM, the adiabatic regime starts for migration
rates slower than $\dot{a}_{Y}=1.0\times10^{-5}$ AU/yr ($D=2.5$
cm), where the calculated probability resembles the theoretical curve
predicted by Henrard's approach (see Fig. \ref{fig3}). For the CM,
the adiabatic regime starts at even slower migration drifts, $\dot{a}_{Y}=2.5\times10^{-7}$
AU/yr ($D=1$ m). This means that in spite of the model, for real
km-size asteroids the Yarkovsky drift should be considered as an adiabatic
regime.

Finally, we observe that in the SEM, and for migration rates of $\dot{a}_{Y}=5.0\times10^{-7}$
AU/yr ($D=0.5$ m) and $\dot{a}_{Y}=2.5\times10^{-7}$ AU/yr ($D=1$
m), besides the tail of delayed orbits, there is a group of orbits
whose final semimajor axes cluster around a value of $a=2.82$ AU.
This corresponds to the 5:2 MMR with Jupiter, which means that even
if these orbits were able to dodge the J3:1 MMR, they did not circumvent
the weaker J5:2 resonance.

\subsection{Integrated Capture Probability\label{intpcap}}

\begin{figure}[ht]
\centering{}\includegraphics[clip,width=0.75\textwidth]{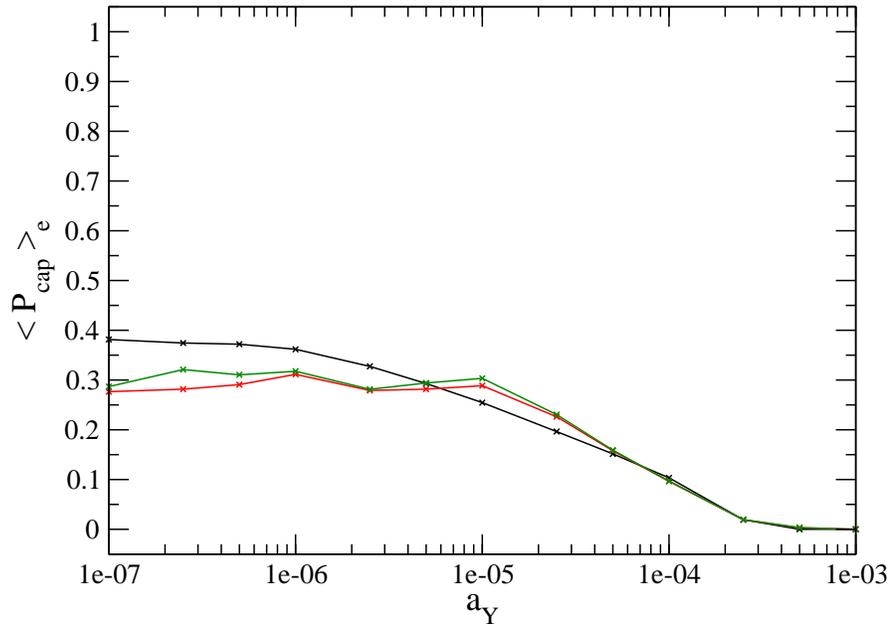}
\caption{Capture probability integrated over the angles and $e$, as a function
of the migration rate. The black curve corresponds to the CM, red
to the EM and green to the SEM.}
\label{fig11}
\end{figure}

In the previous section, we have estimated the capture probability
$P_{cap}(e,\dot{a}_{Y})$ averaged over the angular variables $\theta$
and $\Delta\varpi$. To get an idea of the total capture probability
as a function of the migration rate, we have to integrate the above
estimate over an interval of eccentricity, i.e.: 
\begin{equation}
\langle P_{cap}\rangle_{e}=P_{cap}(\dot{a}_{Y})={\displaystyle \frac{1}{(0.4)}\int_{0}^{0.4}P_{cap}(e,\dot{a}_{Y})\ de.}\label{pcap}
\end{equation}
This is carried out by numerically integrating the curves shown in
Fig. \ref{fig10}. The resulting probability $P_{cap}(\dot{a}_{Y})$
is shown in Fig. \ref{fig11} for the CM (black curve), the EM (green
curve), and the SEM (red curve).

\begin{figure}[ht]
\centering{}\includegraphics[clip,width=1\textwidth]{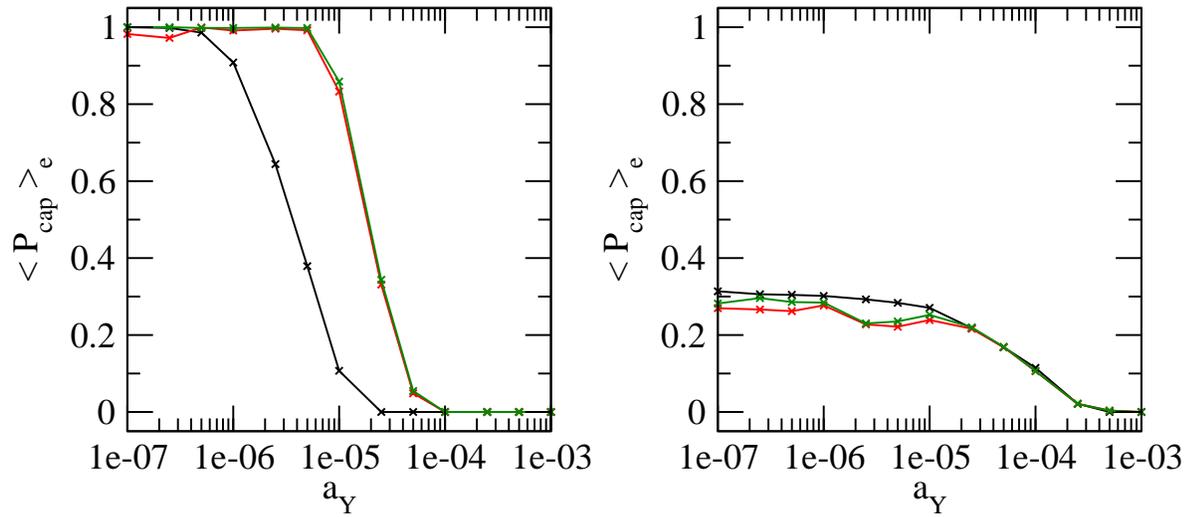}
\caption{Same as Fig. \ref{fig11}, but for two different intervals of $e$.
\emph{Left:} Very low eccentricities. \emph{Right:} Intermediate to
high eccentricities.}
\label{fig12}
\end{figure}

This result indicates that the total capture probability for asteroids
with $0\leq e\leq0.4$ and slow migration rates is of the order of
40 \% in the CM, and 30 \% in the EM and SEM. However, as shown in
Fig. \ref{fig10}, the adiabatic limit occurs for faster migrations
in the EM and SEM rather than in the CM, and should lead to a 100
\% capture probability for $e<e_{c}\approx0.04$. If we divide the
integral Eq. (\ref{pcap}) in two sets, one in the interval $[0,e_{c}]$
and the other in the interval $[e_{c},0.4]$, we obtain the result
shown in Fig. \ref{fig12}. As expected, at very low eccentricities,
the capture probability tend to 100 \% for faster migrations ($\dot{a}_{Y}=5.0\times10^{-6}$
AU/yr; $D=5$ cm) in the more realistic models than in the circular
model ($\dot{a}_{Y}=5.0\times10^{-7}$ AU/yr; $D=50$ cm). On the
other hand, for intermediate/large eccentricities, the capture probability
is independent of the model adopted, and tends to 30 \% in the adiabatic
limit.

Before applying these results to the real case of the Vesta family
and the vestoids, it is interesting to discuss our results with respect
to those obtained by \citet{2006MNRAS.365.1367Q}. The main difference
between Quillen's approach and ours is that she considers a fixed
test particle with a migrating planet (and a migrating resonance),
while we leave the planet fixed and migrate the test particle. In
principle, both approaches should lead to the same result.
However, Quillen did not report in her simulations the existence of
the ``capture windows'' that we observe in our simulations. This
is not surprising, since these structures appear when the initial
angles of the orbits are fixed to the same value, as we showed
in Sect. \ref{dependence}. Since \citet{2006MNRAS.365.1367Q} chose
initial random angles, the windows became hidden in her simulations.
It is worth noting, however, that Quillen's procedure should not affect
the computed capture probabilities, because she also averages the
probabilities over the angular variables. Indeed, the curves shown
in Fig. \ref{fig12} resemble the behavior presented by \citet{2006MNRAS.365.1367Q}
in her figure 3. On the other hand, \citet{2011MNRAS.413..554M}
did detect the capture windows, but they did not provide a clear explanation
for them.

This is a particularly important result, because Quillen's model is
significantly different from ours. Although the basic Hamiltonian
is the same in both models (i.e., an Andoyer-like Hamiltonian), she
includes the non-conservative term directly as a variation of the
mean motion of the planet $n_{1}=n_{1}(t)$, that only affects the
term of the unperturbed Hamiltonian that is linear in the actions.
On the other hand, in our model, we include the non-conservative term
as a repeated ``kick'' in the actions, so it does not only affect
the linear term of the unperturbed Hamiltonian but also the quadratic
term, as well as the amplitudes of the harmonics of the disturbing
function. The fact that two models so different can lead to almost
the same result demonstrates the robustness of the capture process.

\section{Comparison with the Distribution of Real Asteroids\label{compa}}

As a final task, we wish to find the integrated capture probability
for V-type asteroids coming from the Vesta family (vestoids) as a
function of the migration rate. \citet{2008Icar..194..125R} determine
that the probability of a vestoid to cross the J3:1 MMR is of the
order of 3 \%. This result is based on numerical simulations of a
full Solar System model (Venus to Neptune), including real Vesta family
members close to the J3:1 resonance border as massless particles,
and assigning to these asteroids two different drift rates: $\dot{a}_{Y}=1.0\times10^{-10}$
AU/yr ($D=2.5$ km) and $\dot{a}_{Y}=1.0\times10^{-9}$ AU/yr ($D=250$
m).

\begin{figure}[ht]
\centering{}\includegraphics[clip,width=0.75\textwidth]{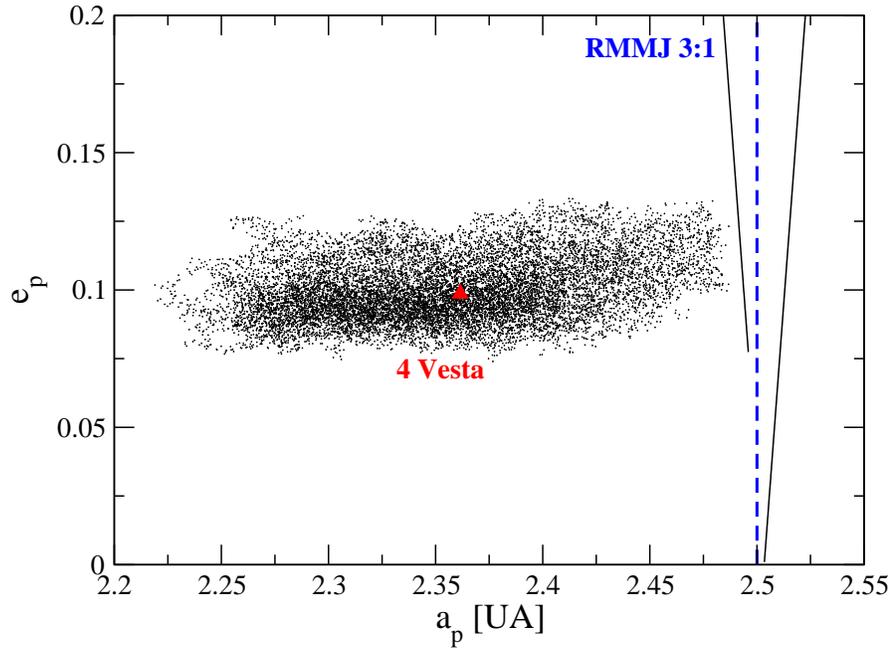}
\caption{The distribution of Vesta family asteroids on the plane of proper
elements $a_{p}-e_{p}$ (black dots). The red triangle indicates the
current location of 4 Vesta. The resonance separatrix (black lines)
and center (blue dashed) are also shown. We are particularly interested
on the evolution of family members close to the resonance border.}
\label{fig13}
\end{figure}

In principle, the Yarkovsky effect should only change the orbital
semimajor axis of a Vesta family member, and thus it is expected that
the Vesta family preserves more or less its original distribution
in eccentricity. The interplay of the Yarkovsky effect with the many
non-linear secular resonances and weak MMRs in inner Main Belt can
produce significant variations of the Vesta family members' eccentricities
(and inclinations) over Gyr time scales (e.g. \citealt{2005A&A...441..819C};
\citealt{2008Icar..193...85N}). However, these variations only affect
a small fraction of the family members and will be disregarded in
our analysis. Figure \ref{fig13} shows the location on the proper
elements plane of the known Vesta family members. Assuming that over
the interval of eccentricities of the Vesta family, that is approximately
$0.07\leq e_{p}\leq0.14$, the proper eccentricity differs very little
from the mean (i.e. averaged over the synodic perturbations) eccentricity,
the integrated capture probability for the family becomes: 
\begin{equation}
\langle P_{cap}\rangle_{e}=P_{Vesta}(\dot{a}_{Y})={\displaystyle \frac{1}{0.07}\int_{0.07}^{0.14}P_{cap}(e,\dot{a}_{Y})\ de.}
\end{equation}

\begin{figure}[ht]
\centering{}\includegraphics[clip,width=0.75\textwidth]{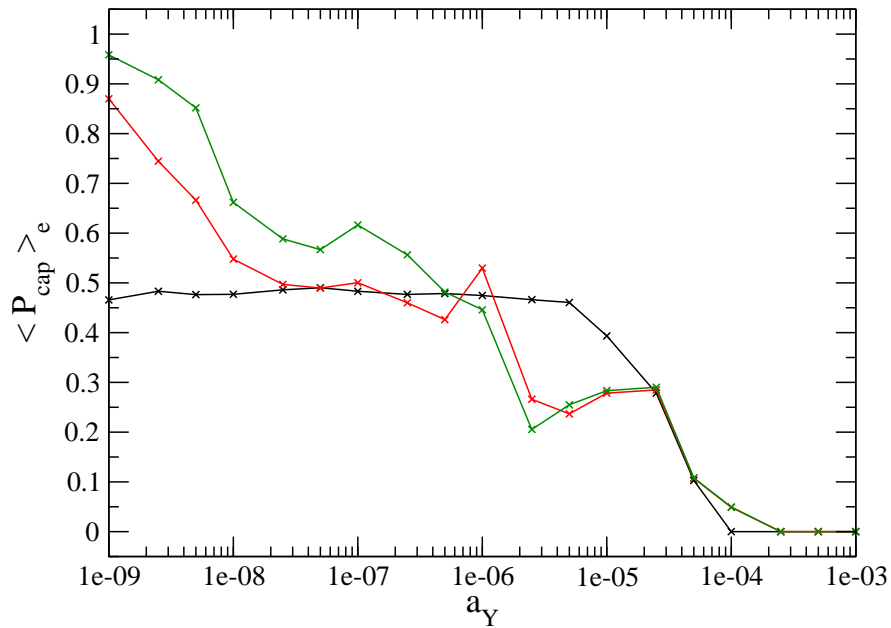}
\caption{Same as Fig. \ref{fig11}, but with the probability integrated between
the eccentricity limits of the Vesta family (0.07 to 0.14). At the
adiabatic limit ($\dot{a}_{Y}<10^{-9}$) both the EM and the SEM predicts
capture probabilities above 90--95 \%, in good agreement with the
results of \citet{2008Icar..194..125R}.}
\label{fig14}
\end{figure}

Figure \ref{fig14} shows the resulting probability for the different
models. In the CM, the probability remains constant around 50 \% for
migration rates slower than $\dot{a}_{Y}=5.0\times10^{-6}$ AU/yr
($D=5$ cm). In the EM, the capture probability reaches 87 \% for
a migration rate of $\dot{a}_{Y}=1.0\times10^{-9}$ AU/yr ($D=250$
m), but the curve shows a growing tendency for smaller rates, and
we may expect to get above 90 \% for drift rates and order of magnitude
slower. The behavior in the SEM is very similar to the EM, but with
higher probabilities that may reach 95--96 \% for the slowest rates
simulated. These results are compatible with those obtained by \citet{2008Icar..194..125R},
and imply that the very low probability that the vestoids have to
cross the J3:1 MMR is basically due to the non-zero eccentricity of
Jupiter. Therefore, it is not necessary to rely on the use of complex
models or time consuming N-body simulations to efficiently reproduce
this behavior.

\section{Conclusions\label{conclu}}

In this work, we have studied the capture vs. crossing probability
into the 3:1 mean motion resonance with Jupiter for an asteroid that
migrates from the inner to the middle Main Belt under the action of
a force that produces a secular change on its orbital semimajor axis.
We were specially interested in the behavior of asteroids belonging
to the Vesta family, that can migrate due to the thermal emission
forces producing the so called Yarkovsky effect.

In order to perform a statistically significant analysis, we developed
an algebraic mapping of the restricted three body problem, averaged
over the synodic angle. The mapping is based on the symplectic approach
developed by \citet{1993CeMDA..56..563H}, but we add the secular
variations on the orbit of the perturber, as well as non-symplectic
terms to simulate the migration. The mapping has the advantage of
being much faster than a full three-body high-order integration, but
keeping the basic features of the behavior of the full model. This
allowed us to perform a huge set of simulations with less computational
cost. Moreover, the mapping model has the advantage that different
parts of the model (eccentricity of Jupiter, secular variations, etc.)
can be switched on and off, thus allowing us to analyze the relevance
of these parts on the actual dynamics.

To simplify our study, we concentrated on three planar models (although
the mapping could be easily extended to take into account the orbital
inclinations of the bodies), according to the behavior of Jupiter's
eccentricity: (i) circular model, (ii) elliptic model, and (iii) elliptic
model with secular variations due to the other Jovian planets. The
mapping results have been compared to numerical simulations of the
full equations of motion for the circular and elliptic models, obtaining
a very good agreement.

At very fast migration rates, most of the asteroids cross the resonance,
while the few that are captured have initial eccentricities within
a given range or ``window''. As the migration rate slows down, this
window shifts to smaller eccentricities and becomes narrower, while
new, even narrower, windows start to appear at higher eccentricities.
At very slow migration rates, the shift of the windows to smaller
eccentricities produces and accumulation of them, and their mutual
overlap generates a region of very low $e$ where the capture probability
is 100 \%, in agreement with the theoretical predictions.

Using the mapping, we have performed simulations of initial conditions
distributed over a line in the $a-e$ plane close to the left branch
of the resonance separatrix. For each initial condition, the initial
angles $\theta=2\sigma$ and $\Delta\varpi$ were distributed between
0 and 2$\pi$. Testing different values of the migration rate, we
arrive to the following results:
\begin{itemize}
\item For the fastest migration rates (i.e. highly non-adiabatic regime)
almost all the asteroids are able to cross the resonance without being
captured. The first captured orbits appear in the elliptic and secular
elliptic models for values of $\dot{a}_{Y}=5.0\times10^{-4}$ AU/yr
($D=0.05$ cm) and slower ones. For the circular model, captures start
at $\dot{a}_{Y}=2.5\times10^{-4}$ AU/yr ($D=0.1$ cm).
\item For the non-adiabatic case, we obtained similar results to those of
\citet{1995CeMDA..61...97G}. The capture probability increases for
increasing eccentricity until it reaches a maximum value (always less
than 1) at an eccentricity $e_{max}$. From this value on, the probability
decreases for increasing eccentricity, tending asymptotically to zero.
Nevertheless, we observe several fluctuations along the probability
curve due to the presence of the above mentioned capture windows.
These fluctuation tend to disappear as we approach the adiabatic case.
\item The limit between the non-adiabatic and adiabatic regimes occurs for
$\dot{a}_{Y}=2.5\times10^{-7}$ AU/yr ($D=1$ m) in the circular model,
and for $\dot{a}_{Y}=5.0\times10^{-6}$ AU/yr ($D=5$ cm) in the elliptic
and secular elliptic models.
\item For both the adiabatic and non-adiabatic regimes, our capture probabilities
show a behavior similar to that described by \citet{2006MNRAS.365.1367Q},
even though her model significantly differs from ours.
\end{itemize}
We computed the total capture probability as a function of the migration
rate, by integrating over a range of eccentricities. Along the range
$0\leq e\leq0.4$, we obtained that, in the adiabatic limit, the probability
tends to 40$\%$ in the circular model and to 30$\%$ in the other
models. Restricting the integral to the range $0\le e\leq0.04$, we
found that the total capture probability is 100$\%$ for a migration
rate of $\dot{a}_{Y}=5.0\times10^{-7}$ AU/yr ($D=50$ cm) in the
circular model, and for $\dot{a}_{Y}=5.0\times10^{-6}$ AU/yr ($D=5$
cm) in the elliptic models. It is worth noting that these rates are
compatible with the rates at which the transition between the non-adiabatic
and adiabatic regimes actually occur. On the other hand, integrating
over the interval $0.04\leq e\leq0.4$, the capture probability tend
to 30 \% in the adiabatic limit, independently of the model. All
these percentages are approximate, and have been estimated from the
outcome of a series of simulations for each system. A complete error
estimation is beyond the scope of this paper, and not necessary for
the current discussion.

Finally, integrating over the range of eccentricities typical of the
Vesta family, $0.07\leq e\leq0.14$, we found that in the circular
model the capture probability tend to 50 \% for $\dot{a}_{Y}\leq5.0\times10^{-6}$
AU/yr (i.e., $D>5$ cm). However, in the elliptic models the probability
is at least 87 \% and 96 \%, respectively, for $\dot{a}_{Y}\leq1.0\times10^{-9}$
AU/yr (corresponding to $D>250$ m). This result is in agreement with
those of \citet{2008Icar..194..125R}, who found that the capture
probability of real asteroids under the perturbation of a full Solar
System model is about 97 \%. We conclude that the high capture probability
of Vesta family members into the J3:1 MMR is basically governed by
the eccentricity of Jupiter and its secular variations. The direct
perturbations of other planets over the asteroids can be disregarded
in the description of this phenomenon.

\begin{acknowledgements}
The authors wish to thank Mira Broz and one anonymous referee for
their comments and suggestions that helped to improve the manuscript.
This work has been supported by CNPq and FAPERJ (Brazil) and 
CONICET (Argentina).
\end{acknowledgements}

\end{document}